\documentclass[aps,showpacs,amsmath,amssymb]{revtex4}

\usepackage{epsfig}

\begin{document}

\title{Light charged particle emission from hot $^{32}$S$^{*}$ formed in $^{20}$Ne + $^{12}$C reaction}

\author {Aparajita Dey\footnote{Electronic address: aparajita.dey@saha.ac.in}\footnote{Present Address: Saha Institute of Nuclear Physics, 1/AF, Bidhan Nagar, Kolkata - 700 064, India.}, S.~Bhattacharya, C.~Bhattacharya, K.~Banerjee, T.~K.~Rana, S.~Kundu, S.~R.~Banerjee, S.~Mukhopadhyay, D.~Gupta\footnote{Present Address: Bose Institute, Dept. of Physics and Centre for Astroparticle Physics and Space Science, Block EN, Sector V, Salt Lake City, Kolkata - 700 091, India.}, R.~Saha }

\affiliation{Variable Energy Cyclotron Centre, Sector - 1, Block - AF,  Bidhan Nagar, Kolkata - 700 064, India. }

\begin{abstract}
Inclusive energy distributions for light charged particles ($p, \  d, \  t$ and $\alpha$) have been measured in the $^{20}$Ne (158, 170, 180, 200 MeV) + $^{12}$C reactions in the angular range 10$^{o}$ -- 50$^{o}$. Exclusive light charged particle energy distribution measurements were also done for the same system at 158 MeV bombarding energy by in-plane light charged particle -- fragment coincidence. Pre-equilibrium components have been separated out from proton energy spectra using moving source model considering two sources. The data have been compared with the predictions of the statistical model code CASCADE. It has been observed that significant deformation effects were needed to be introduced in the compound nucleus in order to explain the shape of the evaporated $d, \ t$ energy spectra. For protons, evaporated energy spectra were rather insensitive to nuclear deformation, though angular distributions could not be explained without deformation. Decay sequence of the hot $^{32}$S nucleus has been investigated through exclusive light charged particle measurements using the $^{20}$Ne (158 MeV) + $^{12}$C reaction. Information on the sequential decay chain has been extracted through comparison of the experimental data with the predictions of the statistical model. It is observed from the present analysis that exclusive light charged particle data may be used as a powerful tool to probe the decay sequence of hot light compound systems.
\end{abstract}

\pacs{ 24.60.Dr, 25.70.Pq, 25.70.Jj, 25.70.Gh}

\maketitle

\section{Introduction}
Light charged particles (LCP) evaporated
from an excited composite is widely used as a tool to investigate the properties of that composite. These emisssions reflect the behavior of the composite at various stages of the deexcitation cascade. In light ion induced reaction the composite system gains low excitation and low angular momentum, while in heavy ion induced reaction the associated angular momentum value is high enough. The statistical model is used to describe the decay of a highly excited composite. The energy spectra of light charged particles emitted in light ion induced reaction are explained satisfactorily by the statistical model predictions \cite{hui89,viesti88}. However, in heavy ion induced reaction, the energy spectra of LCPs are inconsistent with the respective theoretical predictions of statistical model because of the angular momentum induced deformation of the hot composite \cite{awes81}. In addition, specific structure effects ($i.e.$, $\alpha$-clustering in $\alpha$-like nuclei) are also known to influence the reaction process and thus contribute to the deformation of the composite.

The $^{20}$Ne + $^{12}$C system has been investigated extensively in recent years \cite{bhat05,dey06,dey07} in connection with the study of nuclear orbiting in $\alpha$-cluster systems. As orbiting is associated with large deformation of the composite, the deformations of the hot composites formed in the $^{20}$Ne + $^{12}$C reactions (in the energy range 7 -- 10 MeV/nucleon) have been estimated from the study of the respective $\alpha$-evaporation spectra. It was found that the deformations of hot composites are much larger than ``normal'' deformation \cite{dey06}. However, it is interesting to study if the emissions of other light particles are also equally affected by deformation. Here, we report a study on the effect of deformation in the emitted Z = 1 particles, such as, triton, deuteron, proton.

Very recently, there have been a few attempts \cite{dipa01,bhat99,gomez96,shapi97,gomez99} to measure the light charged particles in coincidence with individual evaporation residues to probe the finer details of the reaction meachanism. Such studies may enable us to explore the evaporation decay cascade of the composite system. Moreover, the contributions from other non-evaporative ($i.e.$, fissionlike, pre-equilibrium) decay channels may also be estimated from such measurement. Here, we report an exclusive measurement of light charged particles emitted in coincidence with individual evaporation residues of hot $^{32}$S nucleus produced in the $^{20}$Ne (158 MeV) + $^{12}$C reaction and show that such exclusive data may provide important clues to reveal the intricacies of the decay cascade.

The paper has been arranged as follows. The
experimental procedures are described in the Sec.~II. The analysis 
of the data and the results are discussed in Sec.~III. Finally, the 
summary and concluding remarks are given in Sec.~IV.

\section{Experimental Details}
The experiments were performed with accelerated $^{20}$Ne ion beams
of energies 158, 170, 180 and 200 MeV, respectively, from the
Variable Energy Cyclotron at Kolkata. The target used was $\sim$
550 $\mu$g/cm$^{2}$ self-supporting $^{12}$C. The intermediate mass fragments and
evaporation residues have been detected using two solid state
$\Delta$E -- E [$\sim$ 10 $\mu$m Si(SB) $\Delta$E, $\sim$ 300 $\mu$m
Si(SB) E] telescopes (THI) mounted on one arm of the 91.5 cm
scattering chamber. The light charged particles have been detected
using two solid state thick detector telescopes [$\sim$ 40, 100
$\mu$m Si(SB) $\Delta$E, $\sim$ 5 mm Si(Li) E] (TLI) mounted on the
other arm of the scattering chamber. Typical solid angles subtended
by the telescopes were 0.33 msr, 0.19 msr, 0.74 msr and 0.53 msr,
respectively. The telescopes were calibrated using elastically
scattered $^{20}$Ne ion from Au, Al and C targets and $^{228}$Th
$\alpha$-source. Absolute energy calibrations of the E and $\Delta$E
detectors for each telescopes were done separately using standard
kinematics and energy-loss calculations. The measured energies have
been corrected for the energy losses at the target by incorporating
a single average thickness correction for each fragment energy. The
low-energy cutoffs thus obtained were typically $\sim$ 3.0 MeV for
proton and $\sim$ 10.2 MeV for $\alpha$-particles in TLI telescopes.
In THI telescopes, the cutoffs were $\sim$ 12.3 MeV for Boron and
$\sim$ 19.5 MeV for Oxygen. Well separated bands corresponding to
elements having atomic numbers up to Z = 13 have been identified.

Inclusive energy distributions for light charged particles (Z = 1, 2) emitted in $^{20}$Ne + $^{12}$C reaction have been measured in the angular range 10$^{o}$ to 50$^{o}$, 
in steps of 2.5$^{o}$, at all bombarding energies. LCP energy spectra for $^{20}$Ne + $^{27}$Al at 158 MeV have also been measured for comparison. We have also measured the in-plane coincidence
of light charged particles with intermediate mass fragments and
evaporation residues (ER) for $^{20}$Ne + $^{12}$C at 158 MeV for the
comparison of inclusive and exclusive spectra and to study the decay sequence of hot composite. A part of the present data (inclusive $\alpha$-particle emission) has already been analysed and published elsewhere \cite{dey06}.

\section{Results and Discussions}
Inclusive energy spectra for Z = 1 particles have been analysed for $^{20}$Ne + $^{12}$C reactions at 158, 170, 180 and 200 MeV in the angular range 10$^{o}$ -- 50$^{o}$ to study the effect of nuclear deformation. It had been found from our previous study \cite{dey06} that, the contribution of pre-equilibrium process is significant for proton emission at forward angles only and it becomes negligible at $\theta_{lab}$ $\gtrsim$ 30$^{o}$. Therefore, pre-equilibrium component was extracted from the forward angle ($\theta_{lab} \leq 30^o$) proton data, whereas effect of nuclear deformation was studied using the energy spectra of Z = 1 particles at $\theta_{lab}$ $\gtrsim$ 30$^{o}$, where pre-equilibrium effect is insignificant.

\subsection{Proton emission: Pre-equilibrium processes}

\begin{figure}  [h] \begin{center}
\epsfysize=7.0cm
\epsffile[43 422 434 772]{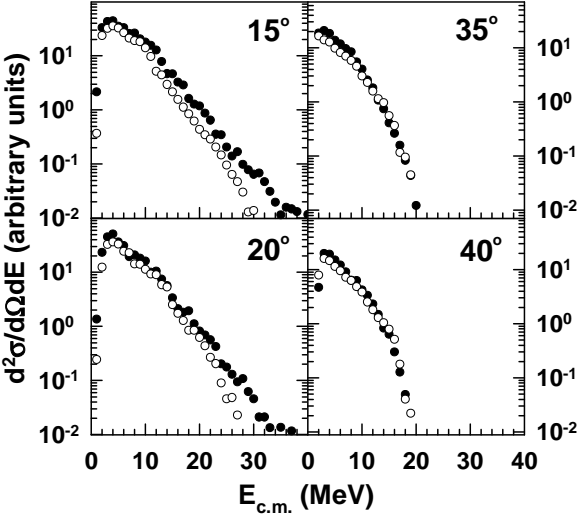}
\end{center}
\caption{Comparison between inclusive ($\bullet$) and exclusive ($\circ$) spectra for protons emitted in $^{20}$Ne (158 MeV) + $^{12}$C at different angles.}
\label{fig1}
\end{figure}

It is evident from Ref.~\cite{dey06} that the inclusive forward angle proton energy distributions have enhanced high-energy `tail' when compared with those at backward angles (Fig.~1 in Ref.~\cite{dey06}). The backward angle spectra is usually believed to have originated from evaporative decay of compound nucleus, and the high energy tails observed in forward direction are considered to be the signature of pre-equilibrium emission. The shapes of the
inclusive proton spectra have been compared with the respective
exclusive spectra (measured in coincidence with ER, Z = 10 -- 13, at
$\theta_{lab}$ = 10$^{o}$) at the angles $\theta_{lab}$ = 15$^{o}$,
20$^{o}$, 35$^{o}$ and 40$^{o}$ in Fig.~\ref{fig1}. It
is found that the shapes of the inclusive (filled circle) and exclusive (open circle) proton
spectra are different for lower angles (15$^{o}$, 20$^{o}$) while,
they are matching well for higher angles (35$^{o}$, 40$^{o}$). This
indicates the presence of pre-equilibrium contributions to the proton spectra at forward
angles. Cross-sections were normalized at
one point for comparison.

\begin{figure}  [h] \begin{center}
\epsfysize=8.0cm
\epsffile[23 205 592 743]{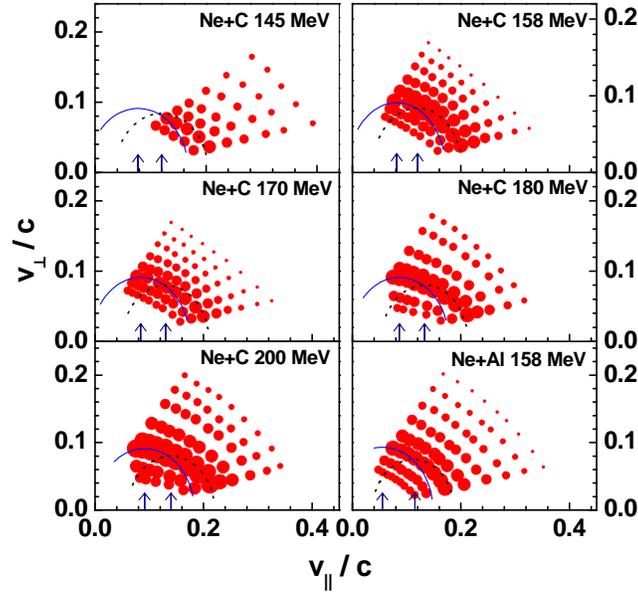}
\end{center}
\caption{(color online) The invariant cross section of protons plotted
in ($v_{\parallel}$,$v_{\perp}$) plane at different energies. The arrows show the equilibrium source velocity (lower) and intermediate source velocity (higher). The most probable average velocities for equilibrium source (solid curve) and intermediate source (dotted curve) are shown for each energy (see text for details).}   
\label{fig2}
\end{figure}

The velocity contour maps of the Galilean-invariant differential
cross-sections, (d$^{2}$$\sigma$/d$\Omega$dE)/pc, as a function of
the velocity of the emitted protons provide an overall picture of the reaction pattern \cite{parker,li88,bel87}. The velocity diagrams of invariant cross-sections in the
($v_{\parallel}$,$v_{\perp}$) plane for protons emitted in $^{20}$Ne + $^{12}$C, $^{27}$Al reactions at different
bombarding energies have been displayed in Fig.~\ref{fig2}. The size of the
point is in proportion to the invariant cross-section. The arrows (lower velocity) indicate the compound nucleus
velocity, $v_{CN}$. The solid circles correspond to the most probable
average velocities of evaporated protons. It has been observed that for all bombarding
energies, the invariant cross-sections do not follow the solid circle.
This further indicates that some component which has higher velocity than
$v_{CN}$ is contributing to the proton emission spectra.

\subsubsection{Extraction of pre-equilibrium component}

From the above discussion it has been found that the slopes of the center-of-mass (c.m.)~energy distributions (Fig.~\ref{fig1}) of emitted protons depend on the emission angles and the invariant cross-section plotted in ($v_{\parallel}$, $v_{\perp}$) plane does not follow the circle centered around $v_{CN}$, the compound nucleus velocity. These observations indicate that the protons are not emitted entirely from a compound nuclear source; emission from other sources, moving with velocity higher than $v_{CN}$, should also be considered to explain the data.

\begin{figure}  \begin{center}
\epsfysize=5.5cm
\epsffile[28 395 563 746]{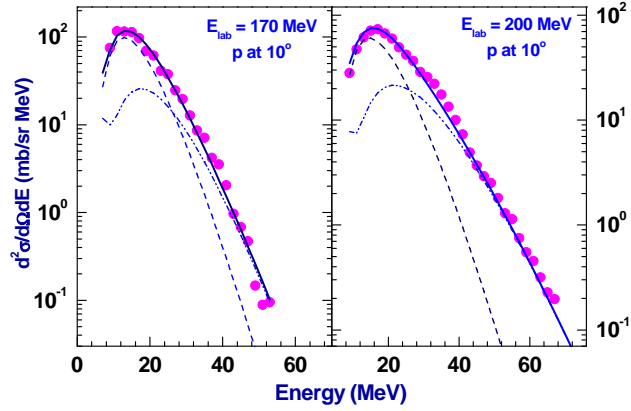} \end{center}
\caption{(color online) The measured proton spectrum emitted in $^{20}$Ne + $^{12}$C reaction at $\theta_{lab}$ = 10$^{o}$ for $E_{lab}$ = 170 and 200 MeV. Experimental data is fitted with Eqn.~\ref{mov2}. The dashed, dash-dot-dotted and solid lines correspond to the equilibrium source ($v_o$, $T_1$), intermediate source ($v_1$, $T_2$) and total contribution, respectively.}
\label{fig3}
\end{figure}

To understand the nature of the sources, the proton energy spectra
have been fitted using a phenomenological moving source model
\cite{li88,bel87,bhat91,bhat93,bord86,bhat95} where it is assumed that the
protons are emitted isotropically in the rest frame of moving
sources. As there is evidence of pre-equilibrium component in proton
spectra, it has been assumed that protons are emitted from two
different sources: (i) equilibrium source having velocity $v_{0}$
and (ii) intermediate velocity source having velocity $v_{1} > v_{0}$. Now
the parametrization expression for double-differential cross section
in laboratory rest frame is
\begin{eqnarray}
\frac{d^2\sigma}{dE_{lab}d\Omega} = C_{1}[(E_{lab} - V_{x1})E_{sx1}]^{3/2} exp(-E_{sx1}/T_{1}) \nonumber \\
+ \  C_{2}[(E_{lab} - V_{x2})E_{sx2}]^{3/2} exp(-E_{sx2}/T_{2}),
\label{mov2}
\end{eqnarray}
where
\begin{eqnarray*}
E_{sx1} = E_{lab} - V_{x1} + E_{0} -2[E_{0}(E_{lab} - V_{x1})]^{1/2}cos\theta_{lab}, \\
E_{sx2} = E_{lab} - V_{x2} + E_{1} -2[E_{1}(E_{lab} - V_{x2})]^{1/2}cos\theta_{lab}.
\end{eqnarray*}
where $C_{1}$ and $C_{2}$ are normalization constants, $E_{1} =
m_{p}v_{1}^2/2$ and $V_{x1}$, $V_{x2}$ are Coulomb correction for the
equilibrium source $v_{0}$ and intermediate velocity source $v_{1}$, respectively. $T_{1}$ and $T_{2}$ are the temperatures of the two sources. In the first step, the energy
spectrum at most backward angle ($i.e.$, at 50$^{o}$ in the present work) has been fitted using a single
source. Then keeping these parameters fixed, the parameters of the
faster moving source ($v_{1}$) have been extracted by two-source
fitting. The optimized values for $v_{o}$, $v_{1}$ and $T_{1}$,
$T_{2}$ are given in Table~\ref{tbl1}. The above procedure is
illustrated in Fig.~\ref{fig3} by dashed line ($C_{1},v_{0},T_{1}$),
dash-dot-dotted line ($C_{2},v_{1},T_{2}$) and solid line (total).
It has been observed that $v_{o}$ for protons are nearly equal to $v_{CN}$. The higher velocity arrows in Fig.~\ref{fig2} correspond to $v_{1}$ and dotted circles represent the average velocities of this type of source.

\begin{figure} 
\begin{center}
\epsfysize=11.5cm
\epsffile[34 76 551 788]{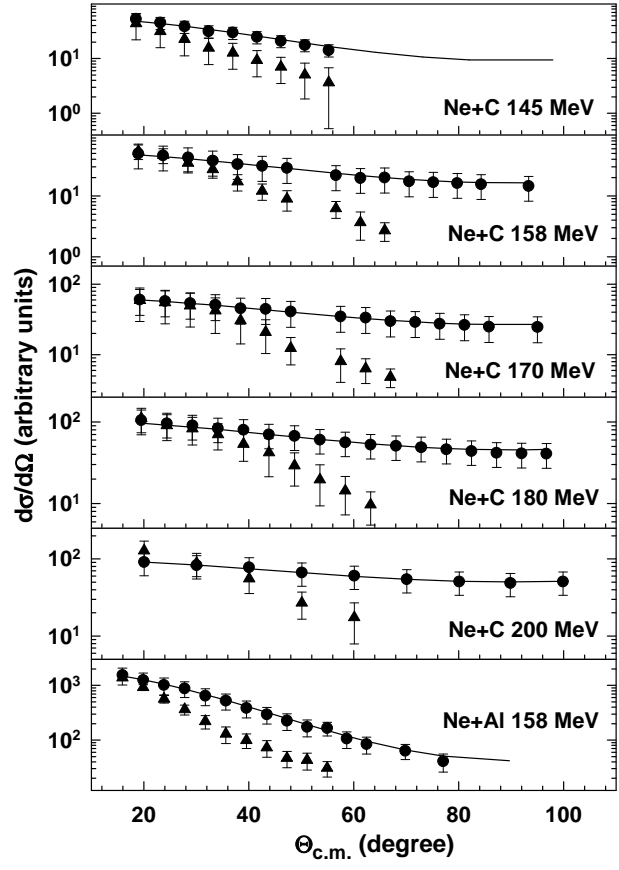}
\end{center}
\caption{The angular distributions for protons emitted in the $^{20}$Ne + $^{12}$C, $^{27}$Al reactions for different bombarding energies. The equilibrium ($\bullet$) and pre-equilibrium ($\blacktriangle$) components are shown. The solid curves are fit to the data using Eq.~\ref{lcp}.}
\label{fig4}
\end{figure}

The center-of-mass angular distributions of the inclusive protons
emitted in $^{20}$Ne + $^{12}$C reactions
have been displayed as a function of the center-of-mass emission
angles in Fig.~\ref{fig4}. The equilibrium (circles) and pre-equilibrium (triangles) components of angular distribution have been shown. It is observed from the figure that for
all bombarding energies, the differential cross-section
for equilibrium component is proportional to $exp(- \beta \sin^{2} \Theta_{c.m.})$, which is the characteristic of the
emission from a thermally equilibrated system \cite{parker}. For a thermally equilibrated system the c.m.~angular distributions are fitted with the following expression:
\begin{equation}
(d\sigma/d\Omega)_{c.m.} = a\; exp(-\beta\; sin^{2}\Theta_{c.m.})
\label{lcp}
\end{equation}
where, $a$ is a normalization constant \\
and
\begin{equation}
\beta = \frac{\hbar^{2} (j + 1/2)^{2}}{2 {\mathcal{I}} T} \frac{\mu R^{2}}{{\mathcal{I}} + \mu R^{2}},
\label{beta}
\end{equation}
$\mathcal{I}$ is moment of inertia, $T$ is temperature, $j$ is angular momentum and $\mu R^{2}$ is the moment of inertia of reduced system. In Fig.~\ref{fig4} the solid lines represent the fits with Eq.~\ref{lcp}. The fitted value of the anisotropy parameter, $\beta_{fitted}$, is given in Table~\ref{tbl2}. The angular distributions for pre-equilibrium component fall off faster than that for equilibrium component as function of angles in case of all bombarding energies. The c.m.~angular distributions for proton emitted in $^{20}$Ne (158 MeV) + $^{27}$Al reaction has also been shown in Fig.~\ref{fig4}.

\subsection{Statistical model analysis using CASCADE}

The analysis has been performed using the standard statistical model code CASCADE \cite{pul77}. In the statistical model calculations, the lower energy part of the LCP spectrum is controlled by the transmission coefficients. The high energy part of the spectrum is controlled by the level density ($E^*,j$) and for a given angular momentum $j$ and excitation energy $E^*$, it is defined as \cite{hui89,pul77}
\begin{eqnarray}
\rho(E, j) = \frac{(2j + 1)}{12} a^{1/2} \left(\frac{\hbar^{2}}{2{\mathcal{I}}_{eff}}\right)^{3/2} \nonumber \\
\times \frac{1}{\left(E + T - \Delta - E_{j}\right)^{2}} exp\left[2 \left[a \left(E - \Delta - E_{j}\right)\right]^{1/2}\right],
\label{eld}
\end{eqnarray}
where $a$ is level density parameter, $T$ is the thermodynamic temperature, $\Delta$ is the pairing correction and $E_j$ is the rotational energy written as 
\begin{equation}
E_{j} = \frac{\hbar^{2}}{2{\mathcal{I}}_{eff}} j(j+1) .
\end{equation}
In hot rotating nuclei, the effective moment of inertia is taken to be spin dependent and is written as
\begin{equation}
{\mathcal{I}}_{eff} = {\mathcal{I}}_{o} \times (1 + \delta_{1}j^{2} + \delta_{2}j^{4}),
\label{ieff}
\end{equation}
where, the rigid-body moment of inertia, $\mathcal{I}$$_{o}$, is given by
\begin{equation}
{\mathcal{I}}_{o} = 2/5 A^{5/3} r_{o}^{2} \ .
\end{equation}
Nonzero values of the deformability parameters $\delta_{1}$ and $\delta_{2}$ introduce the spin dependence in the effective moment of inertia \cite{dey06,dipa02,rkc84}. 

The most sensitive parameters influencing the calculated light charged particle energy spectra are the critical angular momentum $j_{cr}$, the diffuseness parameter $\Delta j$, the level density parameter $a$, the nuclear radius parameter $r_o$, the transmission coefficients $T_j$, and the deformability parameters $\delta_1$ and $\delta_2$.

The transmission coefficients are calculated using the optical model potential; The optical-model parameters for protons, deuterons and tritons were taken from Perey and Perey \cite{pery76} whereas those for $\alpha$-particles were taken from Huizenga and Igo \cite{hui61}. The critical angular momentum for fusion, $j_{cr}$, was calculated using the Bass model \cite{bass} for each bombarding energy. The value of diffusness parameter was taken as 2$\hbar$ for all bombarding energies. 

\subsubsection{Level density parameter}
\begin{figure}  [h]
\epsfysize=10.0cm
\epsffile[20 152 580 794]{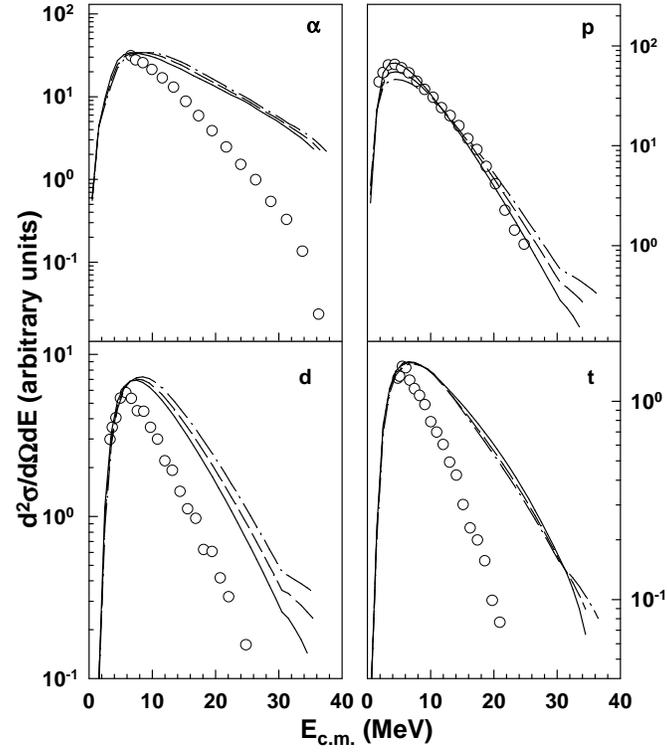}
\caption{Inclusive light charged particle ($\alpha$, $p$, $d$, $t$) energy spectra at $\theta_{lab}$ = 40$^o$ for $^{20}$Ne (200 MeV) + $^{12}$C reaction. The CASCADE predictions with $a$ = $A/8$ (solid),  $A/10$ (dashed), $A/12$ (dash-dotted) are shown.}
\label{fig4a}
\end{figure}

The influence of the level density parameter, $a$, on the energy spectra is displayed in Fig.~\ref{fig4a}. The calculated spectra are shown for three values of $a$: (i) $A/8$ (solid line), (ii) $A/10$ (dashed line), and (iii) $A/12$ (dash-dotted line). The radius parameter $r_o$ was taken to be 1.29 fm, following P\"uhlhofer {\it et al.} \cite{pul77}. The angular momentum dependent deformability parameters ($\delta_1$ and $\delta_2$) were set equal to zero \cite{hui89,dey06,dipa02}. From Fig.~\ref{fig4a}, an increase in yield has been found in the high-energy tail of the light charged particle spectra as the value of the level density parameter is decreased. This is understood in a simple way: It can be seen from Eqn.~\ref{eld} that the level density is exponentially proportional to the square root of $a$. This will reduce the level density when we reduce $a$ and hence energetic particle emission will be favoured. 

It has been found from Fig.~\ref{fig4a} that the theoretical calculations fail to predict the overall
shape of the experimental energy spectra of the light charged particles over the whole energy range except for protons. The value of $a = A/8$ reproduces the proton energy spectrum. Furthermore, the temperature (as well as excitation energy) dependence of $a$ has been investigated recently \cite{mahb04,nebb86,hagel88,gonin90,chbihi91}. The temperature dependence has been incorporated by introducing an inverse level density parameter $K(T)$. Then $a$ becomes 
\begin{equation}
a(T) = A/K(T)
\end{equation}
where $K(T)$ is given by \cite{lest}
\begin{equation}
K(T) = \frac{15.5\ \hbox{MeV}}{1.6 + 1.8 A^{-1/3} - 0.5\{1 - exp[-(\frac{T A^{1/3}}{\hbox{21 MeV}})^2]\}} \ .
\end{equation}
The $K$ value calculated using this equation for $^{20}$Ne + $^{12}$C system is $\sim$ 7.6 MeV in the excitation energy region 74 -- 94 MeV ($E_{lab}$ = 145 -- 200 MeV). Therefore, a value of $a$ = A/8 is reasonable.

\subsubsection{Nuclear radius parameter and deformability parameters}

The most sensitive input parameters are nuclear radius parameters and deformability parameters. Fig.~\ref{fig4a} indicates that the slope of the LCP energy distributions are unexplained by the CASCADE predictions. It is well known fact that increasing the moment of inertia (according to the spin) decreases the slope of the yrast line at higher spin and hence increases the phase space available when emitting light charged particles with lower energies. The spin dependent moment of inertia is given by Eqn.~\ref{ieff}. The slope of the spectra have been well reproduced by varying the parameters $\delta_1$ and $\delta_2$. For different bombarding energies, different sets of $\delta_1$, $\delta_2$ values were needed. The parameter
$\delta_1$ plays the main role in explaining the high-energy part of the data; it is also found to be quite sensitive to variation
of the excitation energy of the system. However, a nonzero value of the parameter $\delta_2$ is required to fit the high-energy-tail
($\gtrsim$ 20 MeV) part of the data in particular, and it is found to be rather insensitive, to the variation of excitation energy, at least
within the range of the present measurements. In addition, the radius parameter, $r_{o}$, was varied from
1.29 fm to 1.35 fm to take care of the low energy side of the experimental spectra. This change of $r_{o}$ causes a shift of the peak of the energy distribution to low energy side and thus helps to fit the lower energy part of the data in a better way. The sensitivity of these nuclear radius parameter and deformability parameters were illustrated in Fig.~5 of Ref.~\cite{dey06} in details. The optimized values of the parameters are given in Table 1 of Ref.~\cite{dey06}.

\subsubsection{Energy distributions of Z = 1 particles}

\begin{figure} [h]
\begin{center}
{\epsfig{file=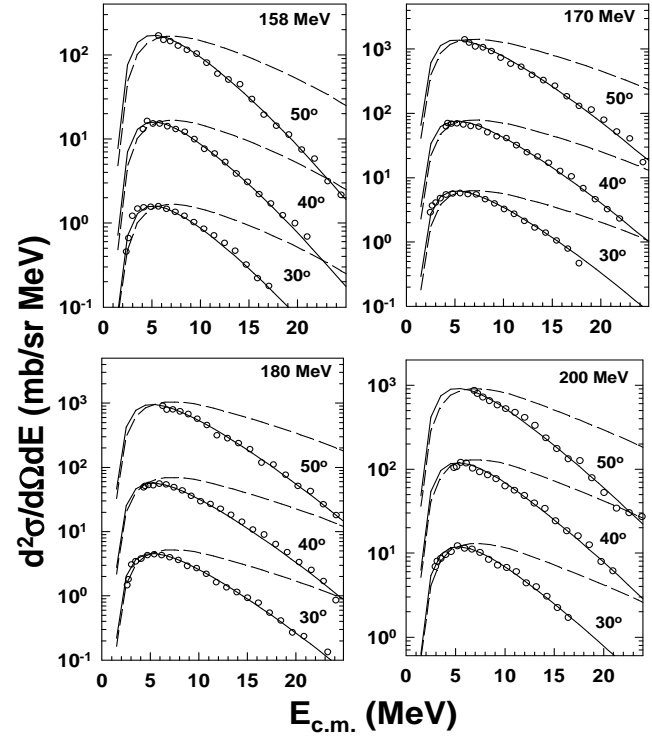,width=8.6cm,height=9.9cm}}
\end{center}
\caption{Inclusive energy spectra for tritons emitted in $^{20}$Ne (158, 170, 180, 200 MeV) + $^{12}$C reactions at angles 30$^{o}$ ($\times$ 10$^{0}$), 40$^{o}$ ($\times$ 10$^{1}$) and 50$^{o}$ ($\times$ 10$^{2}$). Dashed (solid) curves are the predictions of the statistical model code CASCADE without (with) deformation.}
\label{fig5}
\end{figure}

The inclusive energy spectra for triton have been shown in Fig.~\ref{fig5} at $E_{lab}$ = 158, 170, 180 and 200 MeV for few representative laboratory angles. It has been found that the statistical model code CASCADE without deformation parameters ($r_o$ = 1.29 fm, $\delta_1$ = $\delta_2$ = 0) \cite{pul77} could not explain the experimentally measured spectra (dashed line). The CASCADE calculation using the deformation parameters ($r_o$ = 1.35 fm, $\delta_1$ $\neq$ $\delta_2$ $\neq$ 0), which are estimated from $\alpha$-particles spectra \cite{dey06}, explain the data very well (solid line). 

\begin{figure} [h]
\begin{center}
{\epsfig{file=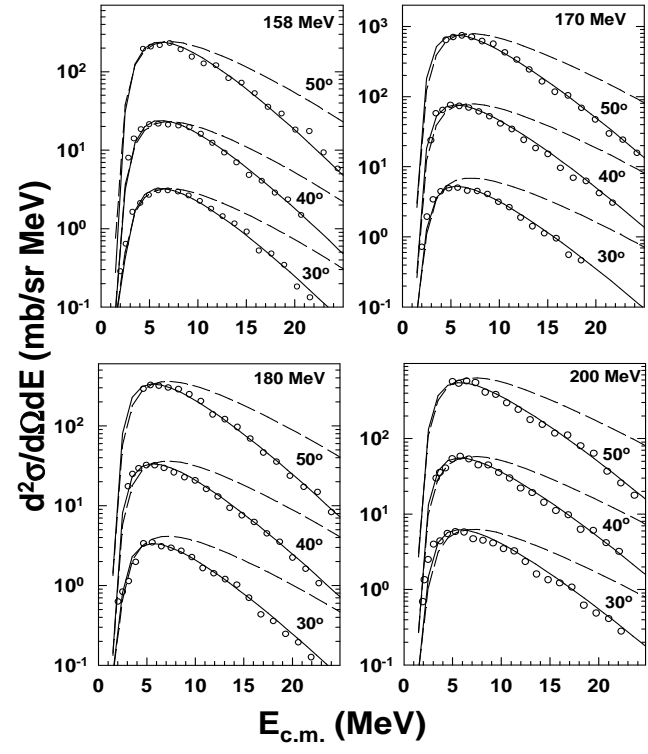,width=8.6cm,height=10.0cm}}
\end{center}
\caption{Same as Fig.~\ref{fig5} for deuterons.}
\label{fig6}
\end{figure}

The inclusive energy spectra for deuteron have been displayed in Fig.~\ref{fig6} at $E_{lab}$ = 158, 170, 180 and 200 MeV for few representative laboratory angles. In this case also, the CASCADE calculations without deformation parameters do not reproduce the data (dashed line). The CASCADE calculations using the deformation parameters (Table 1 in Ref.~\cite{dey06}) estimated from $\alpha$-particle spectra reproduce the data very well (solid line). 

\begin{figure} [h]
\begin{center}
\epsfysize=10.0cm
\epsffile[42 66 560 763]{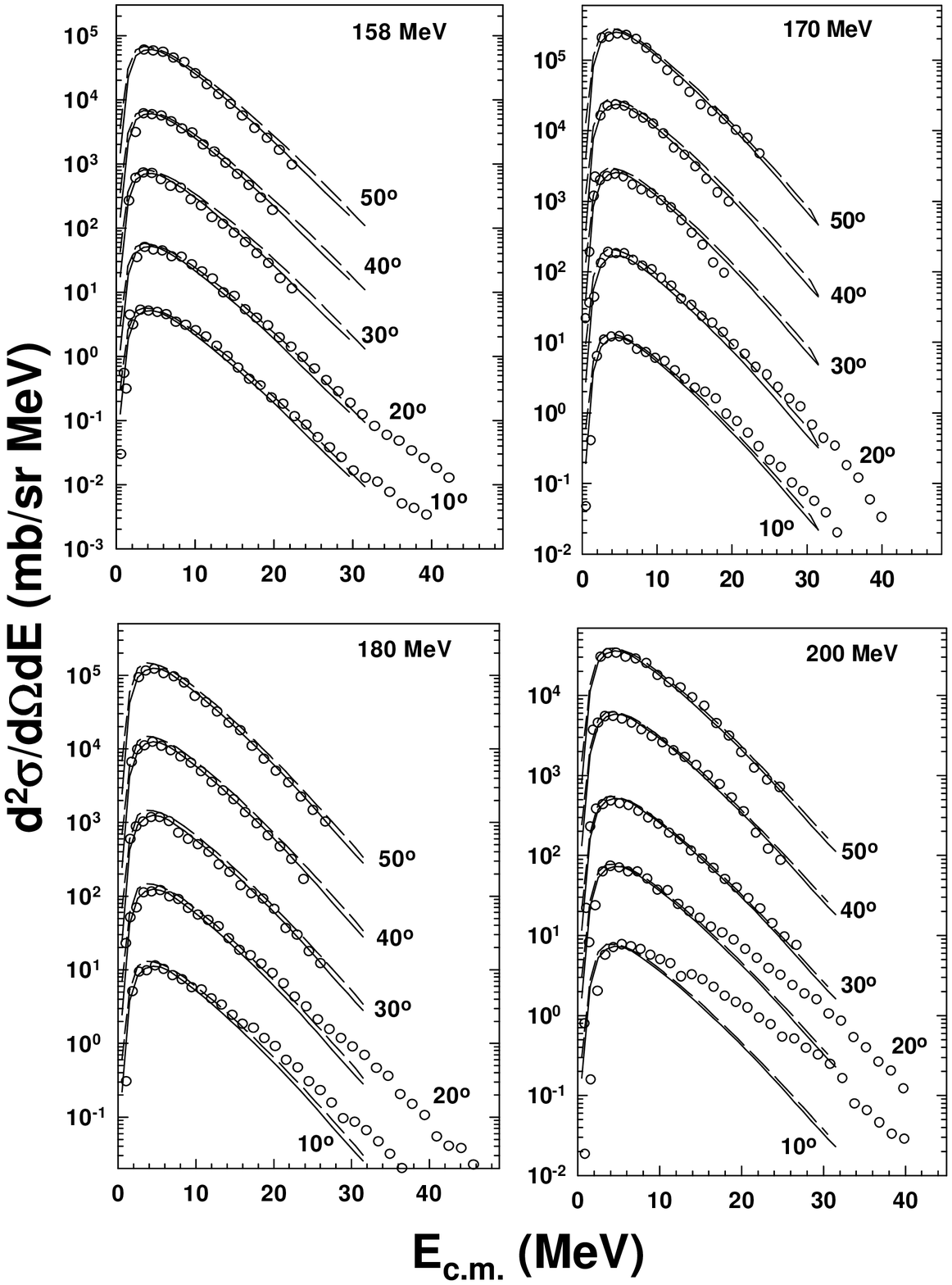}
\end{center}
\caption{Inclusive energy spectra for protons emitted in $^{20}$Ne (158, 170, 180, 200 MeV) + $^{12}$C reactions at the laboratory angles 10$^{o}$ ($\times$ 10$^{0}$), 20$^{o}$ ($\times$ 10$^{1}$), 30$^{o}$
($\times$ 10$^{2}$), 40$^{o}$ ($\times$ 10$^{3}$) and 50$^{o}$ ($\times$ 10$^{4}$). Dashed (solid) curves are the predictions of the statistical model code CASCADE without (with) deformation.}
\label{fig7}
\end{figure}

\begin{figure}  [h] \begin{center}
\epsfysize=8.0cm
\epsffile[32 302 568 803]{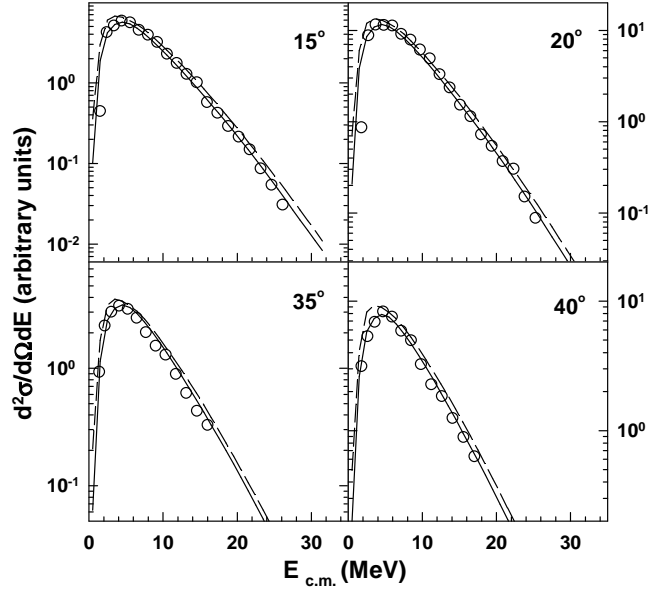}
\end{center}
\caption{Exclusive proton spectra at different
$\theta_{lab}$ measured in coincidence with ERs ($10 \leq Z \leq 13$) emitted at
$\theta_{lab}$ = 10$^{o}$ in $^{20}$Ne (158 MeV) + $^{12}$C reaction. Dashed (solid) curves are the predictions of the statistical model code CASCADE without (with) deformation.}
\label{fig8}
\end{figure}

\begin{figure} [h]  \begin{center}
\epsfysize=6.6cm
\epsffile[40 406 363 760]{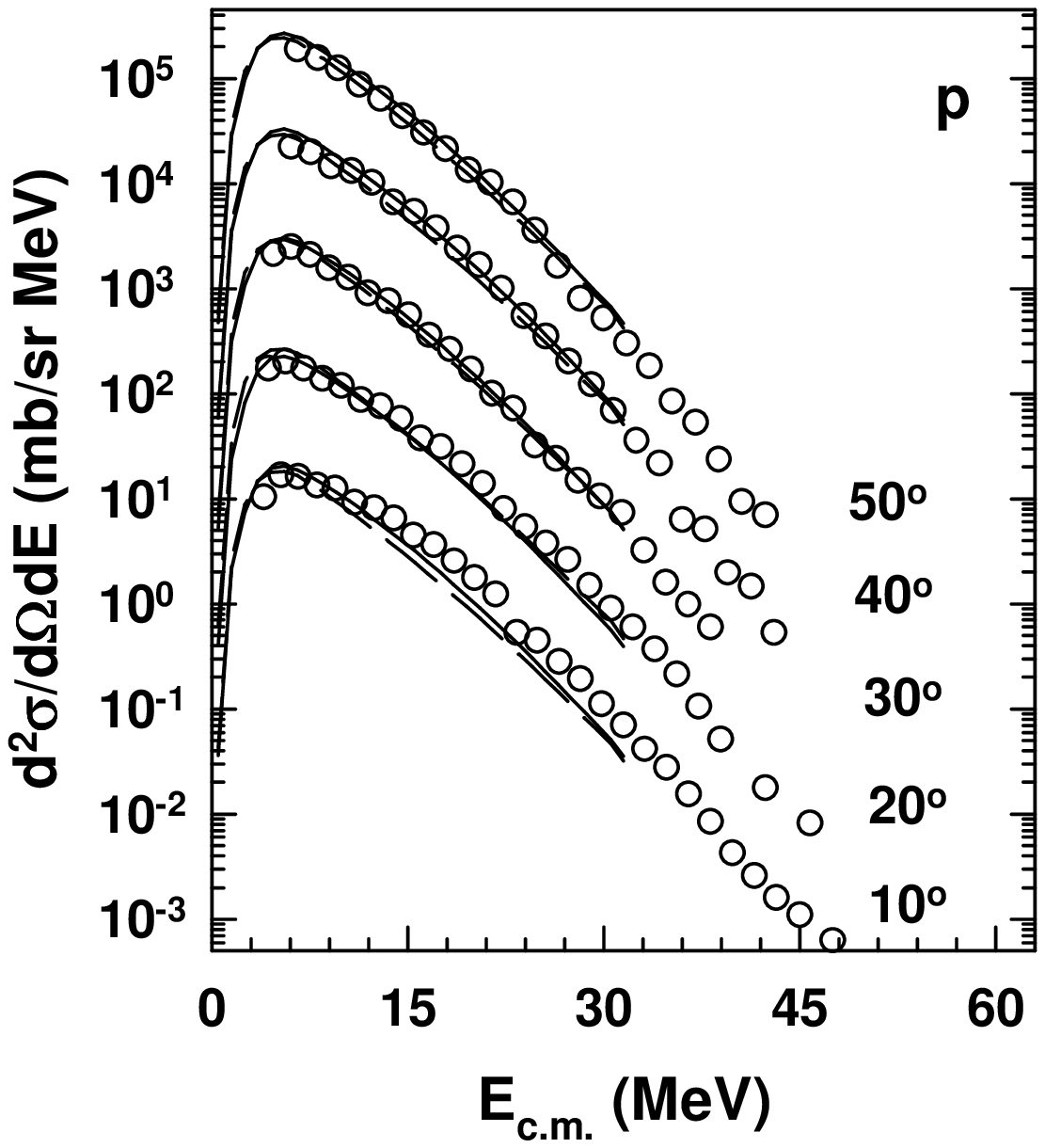}
\end{center}
\caption{Inclusive energy spectra for protons emitted in the $^{20}$Ne (158 MeV) + $^{27}$Al reaction at the laboratory angles 10$^{o}$
($\times$ 10$^{0}$), 20$^{o}$ ($\times$ 10$^{1}$), 30$^{o}$
($\times$ 10$^{2}$), 40$^{o}$ ($\times$ 10$^{3}$) and 50$^{o}$ ($\times$ 10$^{4}$). Dashed (solid) curves are the predictions of the statistical model code CASCADE without (with) deformation.}
\label{fig9}
\end{figure}

The inclusive proton energy spectra are shown in Fig.~\ref{fig7} for different bombarding energies (158, 170, 180 and 200 MeV). It is observed that proton spectra are not sensitive to the deformation of the composite. Dashed line shows the CASCADE calculation without deformation parameters. The calculation with deformation parameters are shown by solid lines. The high energy tail at forward angles is due to the pre-equilibrium processes which are unexplained by the CASCADE calculations. The exclusive proton spectra measured at $E_{lab}$ = 158 MeV in coincidence with evaporation residues (at $\theta_{lab}$ = 10$^{o}$) are shown in Fig.~\ref{fig8}. The shape of these spectra are also insensitive to the deformation. The deformation parameters are $r_o$ = 1.35 fm, $\delta_1$ = 4.5 $\times$ 10$^{-3}$ and $\delta_2$ = 2.0 $\times$ 10$^{-8}$ \cite{dey06}. The protons emitted in the $^{20}$Ne (158 MeV) + $^{27}$Al reaction also show the similar characteristics (Fig.~\ref{fig9}). For this system the deformation parameters are $r_o$ = 1.30 fm, $\delta_1$ = 4.5 $\times$ 10$^{-4}$ and $\delta_2$ = 2.0 $\times$ 10$^{-8}$ and $r_{eff}$ = 1.51 fm \cite{dey06}.

In order to investigate whether the deformation parameters thus obtained can consistently explain the angular distribution data (Fig.~\ref{fig4}), the values of the anisotropy parameter $\beta$ obtained by fitting the angular distribution data using Eqn.~\ref{lcp}, $\beta_{fitted}$, have been compared with those estimated from Eqn.~\ref{beta} using the effective moment of inertia, $\mathcal{I}_{eff}$ (Eqn.~\ref{ieff}). Since $\mathcal{I}_{eff}$ is angular momentum dependent, comparison was done for average angular momentum, $j_{av}$ (= $\frac{2}{3}j_{cr}$). It is interesting to note here that the anisotropy parameters obtained directly by fitting the angular distribution data are in fair agreement with those estimated using the optimized deformation parameters obtained by statistical model fitting of the energy distribution data.

\begin{figure} [h]
\epsfysize=11.0cm
\epsffile[34 50 548 785]{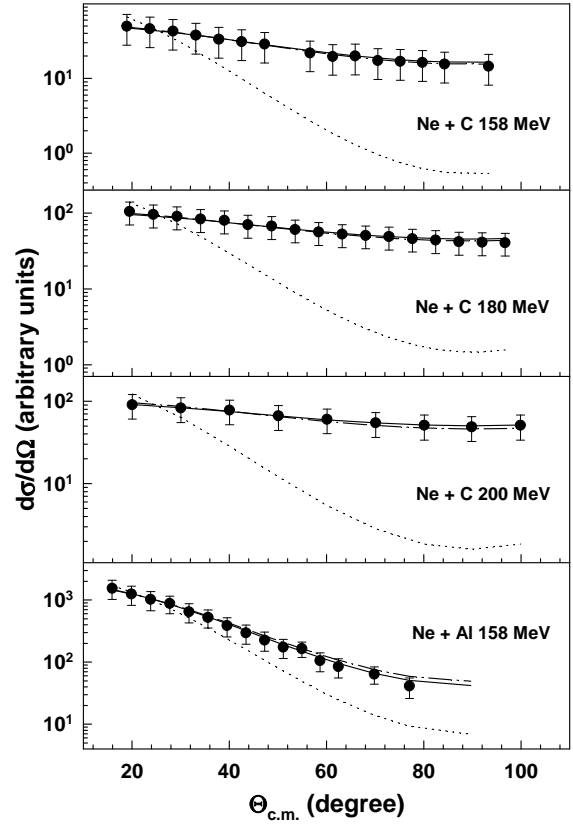}
\caption{The angular distributions for the equilibrium component of protons emitted in the $^{20}$Ne + $^{12}$C, $^{27}$Al reactions for different bombarding energies. The solid curves are fit to the data using Eq.~\ref{lcp}. The dash-dotted and dotted curves correspond to the fit with $\beta_{j_{av}}$ and $\beta_o$, respectively.}
\label{fig9a}
\end{figure}

The $\beta$-values with no deformation ($\beta_o$, obtained using rigid body moment of inertia, $\mathcal{I}_o$) are found to be quite different from $\beta_{fitted}$; so, angular disytribution data cannot be explained properly (Fig.~\ref{fig9a}) with $\beta_o$. It is thus interesting to note that, though the proton energy distribution shapes are rather insensitive to the deformability parameters $\delta_1$, $\delta_2$, it is required to take the deformation into consideration to properly explain the angular distribution data.

\subsection{Prediction of Decay sequence of hot $^{32}$S nucleus}

It is thus evident that the bulk (inclusive) emission data are found to be fairly well explained by `deformed' statistical model predictions. However, it remains to be seen if emission at various stages may also be explained in a similar manner. Exclusive light charged particle emission data may be sensitive to this aspect as indicated earlier \cite{dipa01}. So, the decay sequence of the hot $^{32}$S nucleus has been investigated through the exclusive light charged particle measurements using the $^{20}$Ne (158 MeV) + $^{12}$C reaction. Information on the sequential decay chain has been extracted through comparison of the experimental data with the predictions of the statistical model code CASCADE. 

\begin{figure} [h] \begin{center}
\epsfysize=8.0cm
\epsffile[44 278 552 790]{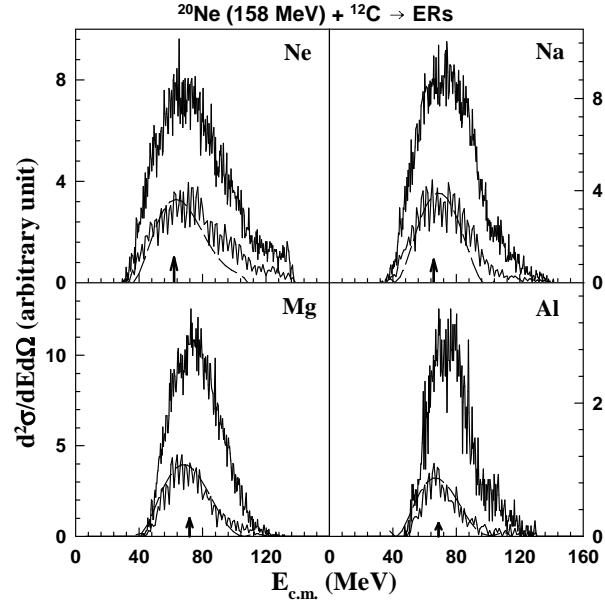} \end{center}
\caption{Inclusive (higher yield) and exclusive (lower yield) energy spectra for Ne, Na, Mg and Al at $\theta_{lab}$ = 10$^{o}$ along with the PACE4 prediction (dashed curve) \cite{pace}. The arrow shows the ER energy corresponding to $v_{CN}\cos\theta_{lab}$.}
\label{fig10}
\end{figure}

Inclusive energy distributions of heavy fragments (spectra having higher yields in Fig.~\ref{fig10}) show that there are significant contribution from fusion-evaporation (FE) component. The arrows in Fig.~\ref{fig10} indicate the evaporation residue (ER) energy corresponding to the ER velocity $v_{CN} \cos\theta_{lab}$. The exclusive energy distributions of the heavy fragments (in coincidence with light charged particles) follow the statistical model prediction (PACE4 \cite{pace}, solid curve in Fig.~\ref{fig10}), which implies that the light charged particles measured in coincidence with Ne, Na, Mg and Al are emitted predominantly through FE channel, in the successive decay of hot $^{32}$S compound nucleus. It is further noted that the heavy fragment spectra are skewed towards higher energy side, which are likely to be due to contributions from the residues of incompletely fused composites (in inverse kinematical systems, velocities of incomplete fusion residues are higher than those of complete fusion residues).

\begin{figure}  [h] \begin{center}
\epsfysize=8.5cm
\epsffile[45 232 503 731]{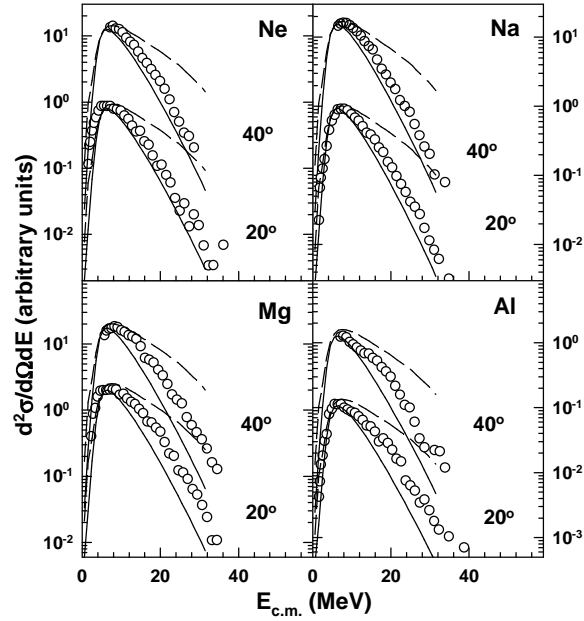}
\end{center}
\caption{Energy distributions of $\alpha$-particles at $\theta_{lab}$ = 20$^{o}$ and 40$^{o}$ measured in coincidence
with Ne, Na, Mg and Al at $\theta_{lab}$ = 10$^{o}$. Solid (dashed) lines are the predictions of code
CASCADE with (without) deformation (see text).}
\label{fig11}
\end{figure}

Some interesting features have been observed in the shapes of the $\alpha$-particle spectra obtained in coincidence with different evaporation residues (Fig.~\ref{fig11}). The solid lines in Fig.~\ref{fig11} represent theoretical predictions of CASCADE \cite{pul77} for the summed evaporation spectra of $\alpha$-particles (sum of all $\alpha$-particle spectra obtained in different decay cascade). A deformed configuration of compound nucleus, represented by the parameters $r_{o}$ = 1.35 fm, $\delta_{1}$ = 4.5 $\times$
10$^{-3}$ and $\delta_{2}$ = 2.0 $\times$ 10$^{-8}$ was used in this calculation. It has been shown in Ref.~\cite{dey06} that statistical model prediction using the above configuration was quite successful in explaining the $\alpha$-particle spectra measured in coincidence with all evaporation residues (Z = 10 -- 13, $\theta_{lab}$ = 10$^{o}$) for the same reaction under consideration. However, the same prescription fails to explain the observed $\alpha$-particle spectra (open circles) in coincidence with individual evaporation residues as evident from Fig.~\ref{fig11}. It is indicative of the fact that the $\alpha$-particles follow some specific decay path to populate a particular evaporation residue and the path is different as one goes from one residue to another. Thus, the study of light particle spectra in coincidence with individual residues is likely to reveal some interesting details of the compound nuclear decay sequence.

In order to understand the shapes of the exclusive $\alpha$-particle energy spectra observed in coincidence with individual residue, $\alpha$-particle evaporation spectra from each possible stage of the decay cascade populating a particular residue have been calculated and summed up with appropriate weightage to generate exclusive $\alpha$-particle spectra corresponding to each residue; these spectra have then been compared with the respective experimental data. The procedure is illustrated as follows: Production of a typical evaporation residue, $e.g.$, Al, is possible through one of the following decay paths,
\begin{subequations}
\begin{eqnarray} 
	 ^{32}S \stackrel{\alpha}\longrightarrow \  ^{28}Si \stackrel{p}\longrightarrow \  ^{27}Al  \stackrel{n}\longrightarrow \  ^{26}Al ~~~~ \label{subeq:1}  \\
	 ^{32}S \stackrel{p}\longrightarrow \  ^{31}P  \stackrel{\alpha}\longrightarrow \  ^{27}Al \stackrel{n}\longrightarrow \  ^{26}Al ~~~~  \label{subeq:2} \\
	 ^{32}S \stackrel{n}\longrightarrow \  ^{31}S \stackrel{\alpha}\longrightarrow \  ^{27}Si \stackrel{p}\longrightarrow \  ^{26}Al ~~~~  \label{subeq:3}  \\
	 ^{32}S \stackrel{n}\longrightarrow \  ^{31}S \stackrel{p}\longrightarrow \  ^{30}P \stackrel{\alpha}\longrightarrow \  ^{26}Al ~~~~ \label{subeq:4}   \\
	 ^{32}S \stackrel{n}\longrightarrow \  ^{31}S \stackrel{p}\longrightarrow \  ^{30}P \stackrel{n}\longrightarrow \   ^{29}P \stackrel{\alpha}\longrightarrow \   ^{25}Al ~~~~  \label{subeq:5}  \\
	 ^{32}S \stackrel{p}\longrightarrow \  ^{31}P \stackrel{p}\longrightarrow \  ^{30}Si \stackrel{p}\longrightarrow \  ^{29}Al ~~~~ \label{subeq:6} \\
	 ^{32}S \stackrel{p}\longrightarrow \, ^{31}P \stackrel{n}\longrightarrow \, ^{30}P \stackrel{p}\longrightarrow \, ^{29}Si \stackrel{n}\longrightarrow \, ^{28}Si \stackrel{p}\longrightarrow \, ^{27}Al ~~~~  \label{subeq:7}
\end{eqnarray}
\end{subequations}

\begin{figure} [h] \begin{center}
\epsfysize=10.0cm
\epsffile[30 70 560 719]{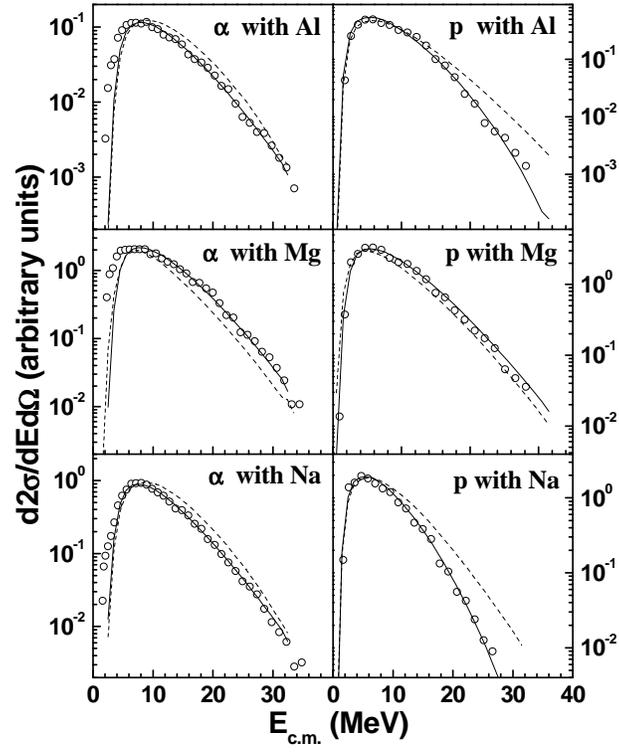} \end{center}
\caption{$\alpha$-particles and protons energy spectra measured at $\theta_{lab}$ = 20$^o$ in coincidence with individual residue. Statistical model spectra are shown by dashed (sum of all possible decay path) and solid (particular decay path) curves (see text).}
\label{fig12}
\end{figure}

\noindent
So, to generate the $\alpha$-particle spectrum in coincidence with Al, the contribution of each decay path (\ref{subeq:1} -- \ref{subeq:7}) is added with the appropriate weightage factor (yield of decay path / total yield of all decay paths). This CASCADE prediction is shown by dashed line in Fig.~\ref{fig12}. The similar procedure has been done for the proton spectra. It is interesting to find that the exclusive $\alpha$-particle and proton spectra are still unexplained by the CASCADE predictions. The spectra measured in-coincidence with Mg and Na are also unexplained by the spectra calculated using this technique (see Fig.~\ref{fig12}).

\begin{figure} [h] \begin{center}
\epsfysize=10.0cm
\epsffile[35 189 536 801]{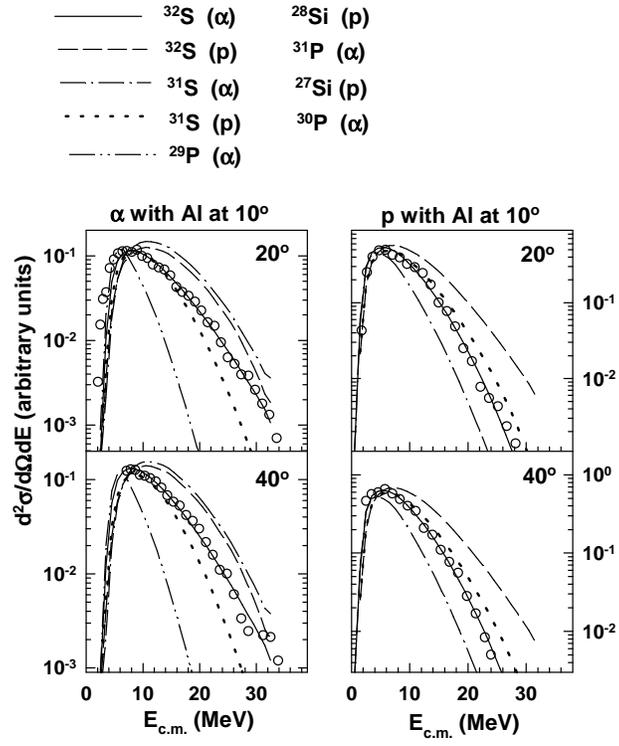} \end{center}
\caption{Energy distributions of $\alpha$-particles and protons measured in coincidence
with Al at different laboratory angles. Different curves are
the statistical model predictions using code CASCADE for the emission
of $\alpha$-particles and protons from different nuclei in the decay chain (see
text).}
\label{fig13}
\end{figure}

The $\alpha$-particles measured in coincidence with Al may be emitted from one or more of the following parent nuclei: $^{32}$S, $^{31}$S, $^{31}$P, $^{30}$P, $^{29}$P. The experimentally measured $\alpha$-particle spectra (open circle) in coincidence with Al have been displayed in Fig.~\ref{fig13}. The theoretical $\alpha$-particle spectra calculated from different stages of the decay cascade ($i.e.$, $^{32}$S, $^{31}$P, $^{31}$S, $^{30}$P, and $^{29}$P) using CASCADE have also been displayed in Fig.~\ref{fig13} (represented by solid, dashed, dash-dotted, dotted and dash-dot-dotted lines, respectively). It is evident from Fig.~\ref{fig13} that the calculated first chance emission of $\alpha$-particle from $^{32}$S nucleus agrees well with the measured spectral shape. This is further supported by the proton spectra measured in coincidence with Al, which clearly indicate that the residue Al may be formed predominantly by the decay of $^{32}$S nucleus through the emission of $\alpha$-particle followed by a proton (path \ref{subeq:1}), $i.e.$,
\begin{center}
$^{32}S  \stackrel{\alpha}{\longrightarrow} \  ^{28}Si \stackrel{p}{\longrightarrow} \  ^{27}Al$,
\end{center}
though relative $\alpha$ emission probabilities are comparable for both \ref{subeq:1} and \ref{subeq:2} paths.

\begin{figure} [h] \begin{center}
\epsfysize=10.0cm
\epsffile[41 189 530 766]{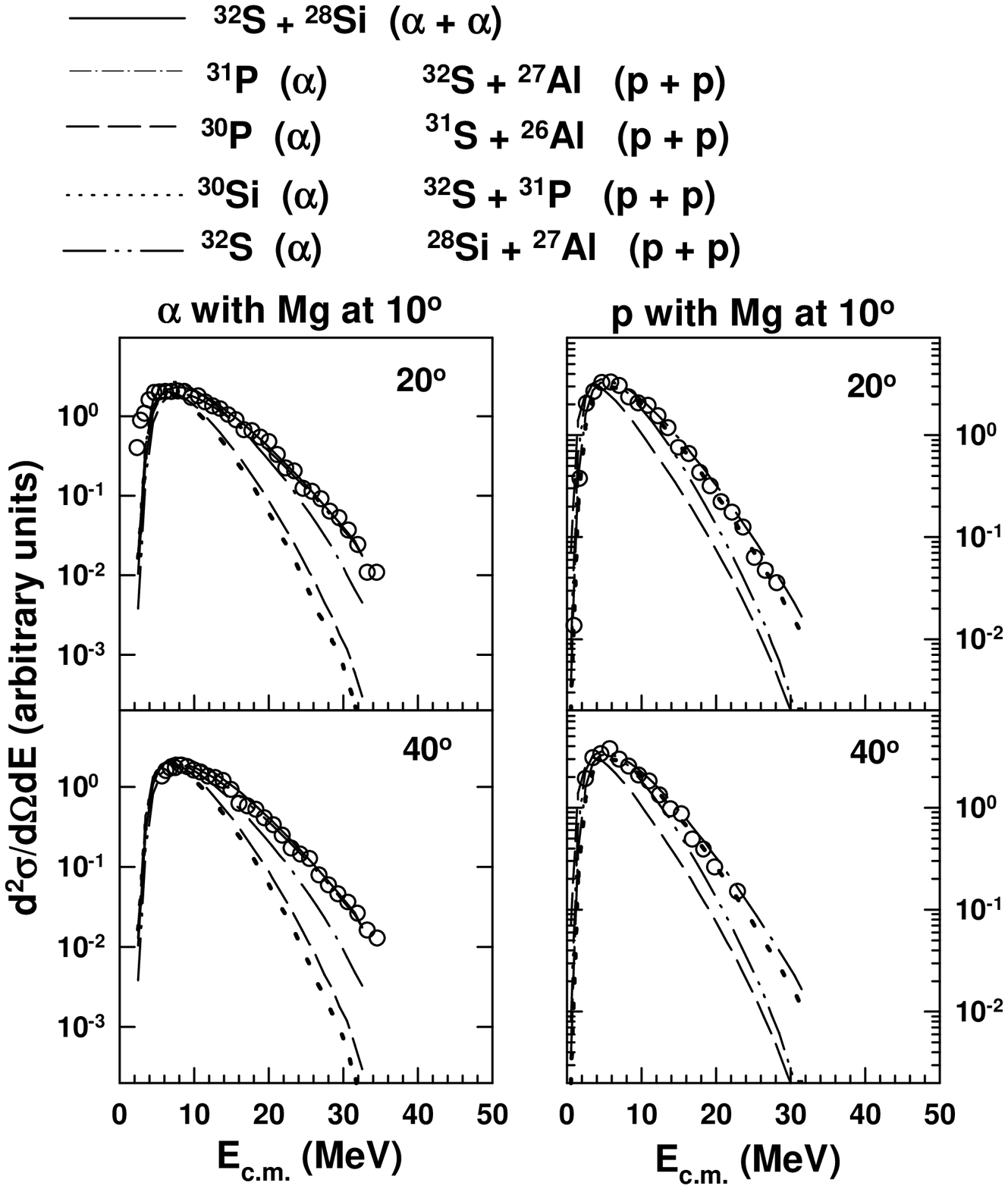} \end{center}
\caption{Same as Fig.~\ref{fig13} for $\alpha$-particle and p in coincidence with Mg.}
\label{fig14}
\end{figure}

The possible parent nuclei which may emit $\alpha$-particles to
populate Mg are $^{32}$S, $^{31}$P, $^{30}$P, $^{30}$Si and $^{28}$Si. The
measured $\alpha$-particle spectra (open circles) in coincidence
with Mg for different laboratory angles have been displayed in
Fig.~\ref{fig14} along with the respective CASCADE predictions.
The solid line corresponds to the total $\alpha$-particle spectra obtained from the CASCADE
calculation for the decay sequence: 
\begin{equation}
 ^{32}S \stackrel{\alpha}{\longrightarrow} \  ^{28}Si \stackrel{\alpha}{\longrightarrow} \  ^{24}Mg.
 \label{aa}
\end{equation}
It is observed that this decay path reproduce the experimental spectra very well. The dash-dotted, dashed, dotted and dash-dot-dotted lines show the theoretical (CASCADE prediction) $\alpha$-particle spectra for the nuclei $^{32}$S, $^{31}$P, $^{30}$P and $^{30}$Si, respectively. 

The possible parent nuclei which may emit proton to populate Mg are $^{32}$S, $^{31}$S, $^{31}$P, $^{28}$Si, $^{27}$Si, $^{27}$Al and $^{26}$Al. The possible decay sequences are
\begin{subequations}
\begin{eqnarray} 
	^{32}S \stackrel{p}{\longrightarrow} \  ^{31}P \stackrel{\alpha}{\longrightarrow} \  ^{27}Al \stackrel{p}{\longrightarrow} \  ^{26}Mg  \label{subeq:a} \\
	^{32}S \stackrel{n}{\longrightarrow} \  ^{31}S \stackrel{p}{\longrightarrow} \  ^{30}P \stackrel{\alpha}{\longrightarrow} \  ^{26}Al \stackrel{p}{\longrightarrow} \  ^{25}Mg  \label{subeq:b} \\
	^{32}S \stackrel{p}{\longrightarrow} \  ^{31}P \stackrel{p}{\longrightarrow} \  ^{30}Si \stackrel{\alpha}{\longrightarrow} \  ^{26}Mg   \label{subeq:c} \\
	^{32}S \stackrel{\alpha}{\longrightarrow} \  ^{28}Si \stackrel{p}{\longrightarrow} \  ^{27}Al \stackrel{p}{\longrightarrow} \  ^{26}Mg \label{subeq:d} \\
	^{32}S  \stackrel{n}{\longrightarrow} \  ^{31}S \stackrel{\alpha}{\longrightarrow} \  ^{27}Si \stackrel{p}{\longrightarrow} \  ^{26}Al \stackrel{p}{\longrightarrow} \  ^{25}Mg   \label{subeq:e}
\end{eqnarray}
\end{subequations}
From Fig.~\ref{fig14}, it is seen that the shapes of the experimental $\alpha$-particle spectra are in fair agreement with the spectra obtained from decay path \ref{subeq:d} (coincident with $\alpha$-spectra from path \ref{aa}), however, the experimental proton energy distributions match well with those of theoretical predictions for the case of proton emission from decay paths~\ref{subeq:a}~and~\ref{subeq:c}. The $\alpha$-particle spectra obtained from path~\ref{subeq:c} are very much different from the experimental spectra. The $\alpha$-particle spectra from path~\ref{subeq:a} and the proton spectra from path~\ref{subeq:d} are in well agreement with lower energy part of the respective experimental spectra, though the higher energy parts remain unexplained. Therefore, from the analyses of exclusive proton and $\alpha$-particle spectra measured in coincidence with Mg, it may be inferred that Mg may have been populated in the decay of $^{32}$S nucleus through both $p - \alpha - p$ emission (decay path~\ref{subeq:a}) and $\alpha - p - p$ emission (decay path~\ref{subeq:d}), in addition to the exclusive $2\alpha$ decay route (path \ref{aa}). According to statistical model, the probability of different decay paths, through which Mg can be populated, decrease in the order \ref{subeq:a}, \ref{subeq:c}~and~\ref{subeq:d}, and \ref{aa}.

\begin{figure} [h] \begin{center}
\epsfysize=9.8cm
\epsffile[35 218 530 778]{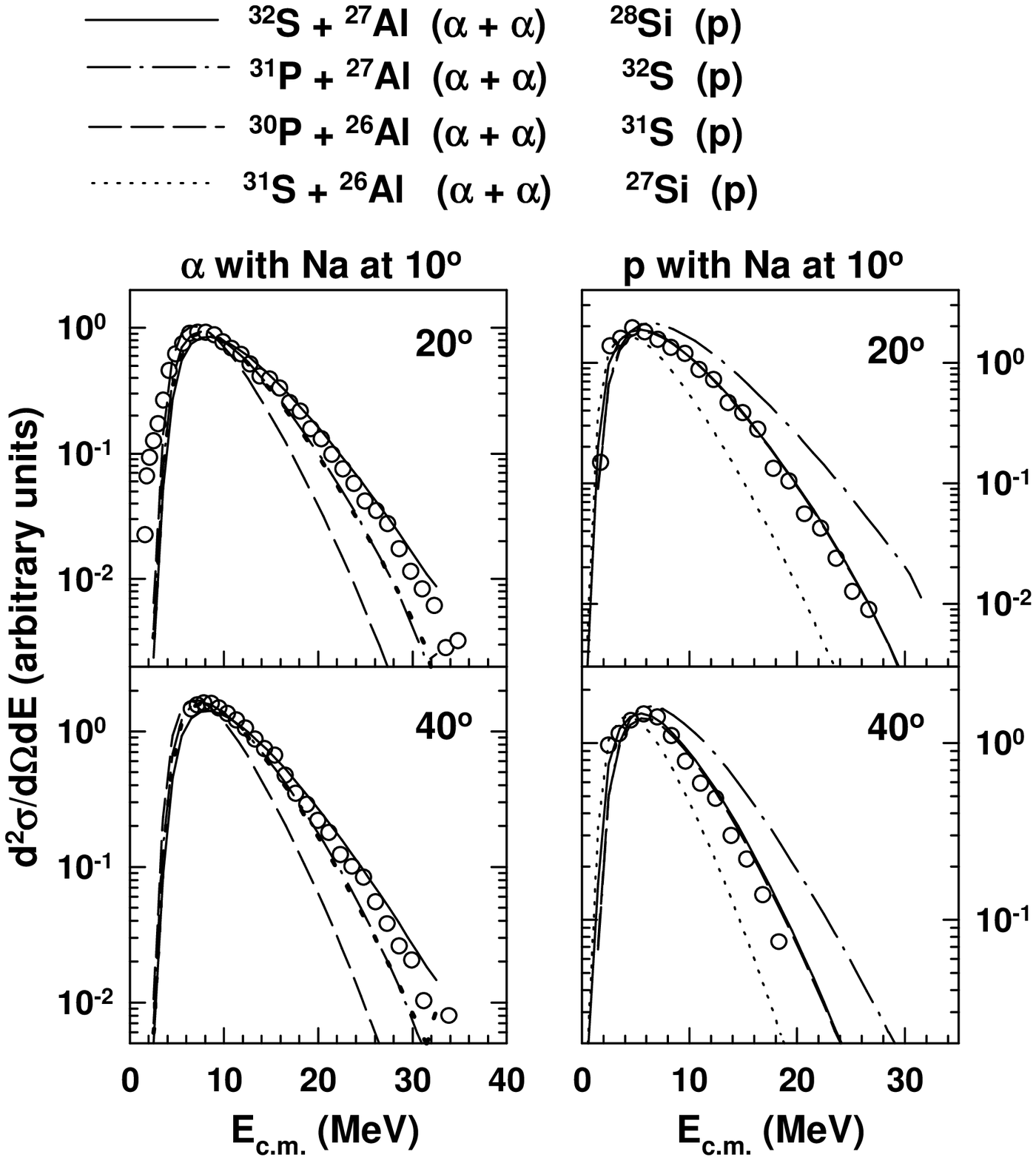}
\end{center}
\caption{Same as Fig.~\ref{fig13} for $\alpha$-particle and p in coincidence with Na.}
\label{fig15}
\end{figure}

The $\alpha$-particle and proton energy spectra measured in coincidence with the residue Na have been displayed in Fig.~\ref{fig15}. The parent nuclei which may populate Na through $\alpha$-particle emission are $^{32}$S, $^{31}$S, $^{31}$P, $^{30}$P, $^{27}$Al and $^{26}$Al. The protons may be emitted from $^{32}$S, $^{31}$S, $^{28}$Si, $^{27}$Si, $^{27}$Al, $^{26}$Mg nuclei. The different decay paths are
\begin{subequations}
\begin{eqnarray}
^{32}S \stackrel{\alpha}{\longrightarrow} \  ^{28}Si \stackrel{p}{\longrightarrow} \  ^{27}Al \stackrel{\alpha}{\longrightarrow} \  ^{23}Na  \label{na1}  \\
^{32}S \stackrel{p}{\longrightarrow} \  ^{31}P \stackrel{\alpha}{\longrightarrow} \  ^{27}Al \stackrel{\alpha}{\longrightarrow} \  ^{23}Na  \label{na2}  \\
^{32}S \stackrel{n}{\longrightarrow} \  ^{31}S \stackrel{\alpha}{\longrightarrow} \  ^{27}Si \stackrel{p}{\longrightarrow} \  ^{26}Al \stackrel{\alpha}{\longrightarrow} \  ^{22}Na   \label{na3}  \\
^{32}S \stackrel{n}{\longrightarrow} \  ^{31}S \stackrel{p}{\longrightarrow} \  ^{30}P \stackrel{\alpha}{\longrightarrow} \  ^{26}Al \stackrel{\alpha}{\longrightarrow} \  ^{22}Na  \label{na4}  \\
^{32}S \stackrel{\alpha}{\longrightarrow} \  ^{28}Si \stackrel{p}{\longrightarrow} \  ^{27}Al \stackrel{p}{\longrightarrow} \  ^{26}Mg \stackrel{p}{\longrightarrow} \  ^{25}Na  \label{na5}
\end{eqnarray}
\end{subequations}

From Fig.~\ref{fig15} it is evident that the residue Na may have been populated predominantly through the sequence \ref{na1} (solid line). However, for the residue Na, \ref{na1} and \ref{na2} (dash-dotted line) decay paths have more or less equal probability. The other decay paths \ref{na3} (dotted line) and \ref{na4} (dashed line) have nearly same probabilities, which are five times smaller than the earlier two paths (\ref{na1} and \ref{na2}). 

Therefore, from this analysis it has been found that population of evaporation residue follows a specific decay path (solid lines in Fig.~\ref{fig12}).  

\section{Summary}
In this work, the $^{20}$Ne + $^{12}$C system has been studied at four incident energies, 158, 170, 180, and 200 MeV, respectively. The inclusive energy spectra of proton, deuteron and triton have been measured at different laboratory angles in the range 10$^o$ to 50$^o$. It was found that the invariant cross-section for protons plotted in $(v_{\parallel} - v_{\perp})$ plane does not follow the circle centered around compound nucleus velocity. This indicates the presence of a source which has velocity higher than $v_{CN}$. The effect of pre-equilibrium process has been found in the emitted proton spectra at forward angles ($\theta_{lab} \leq 30^o$). This pre-equilibrium component has been extracted using the moving source model considering two sources, equilibrium source and intermediate velocity source. The $^{20}$Ne (158 MeV) + $^{27}$Al reaction has been studied for comparison, and this also shows the presence of pre-equilibrium process.

It has been observed that nuclear deformation affected the energy distributions of different Z = 1 particles in different manner. The proton energy spectra were found to be rather insensitive to nuclear deformation for both the systems ($^{20}$Ne + $^{12}$C, $^{27}$Al) studied here. On the other hand, deuteron and triton energy spectra could not be explained without the introduction of deformation; in both cases ($d$ and $t$), it was found that same set of spin dependent deformability parameters ($\delta_1$, $\delta_2$) which was required to fit the alpha-particle spectra \cite{dey06}, can explain the data fairly well. It is interesting to note here that for protons, though the the energy distributions are rather insensitive to nuclear deformation, proper explanation of the angular distributions of evaporated protons would require the deformation to be incorporated in the calculation, and the magnitudes of the deformation parameters were same as those used for fitting the $d, \ t$ and $\alpha$ energy spectra. The sensitivity of the statistical model prediction of the energy spectra on  level density parameter $a$  has also been studied and it was found that the value of $a=A/8$ generated best fit to the data.

Some interesting features in the shapes of the exclusive light charged particle spectra have been observed in the decay of $^{32}$S nucleus which may be correlated, at least in a qualitative manner, with the decay sequence of hot composite system. It may be inferred from the $\alpha$-particle and proton spectra observed in-coincidence
with Al that, the decay sequence in this case is first chance $\alpha$ emission followed by a proton emission. Similarly, in case of Mg, it may be inferred that Mg
has been populated by sequential emission of $\alpha$-particles from $^{32}$S nucleus. In addition, Mg may also populate through $p - \alpha - p$ and $\alpha - p - p$ emission paths. For Na residue, $\alpha - p - \alpha$ is the most prominent decay path. 

It is thus clear that, the exclusive fragment-light particle data may be used to extract information about the nuclear decay sequence. However, it is not intuitively evident why some particular decay paths would be favoured against other statistically competitive paths to populate a particular residue. More detailed exclusive measurements are required to have a concrete idea about the decay sequence. In addition, refinement of statistical model codes is also needed to tract down the complete decay sequence in more unambiguous manner.

\acknowledgements

The authors would like to thank the cyclotron operating staff for smooth running of the machine during the experiment and H. P. Sil for the fabrication of thin silicon detectors for the experiment.

\newpage

Tables

\begin{table}  [h]
\caption{Optimized parameters from moving source model for
protons.}
\begin{center}
\begin{tabular}{cc|ccccc} \hline \hline
$E_{lab}$&$v_{CN}/c$&$\theta_{lab}$&$v_{0}/c$&$T_{1}$&$v_{1}/c$&$T_{2}$ \\
(MeV)&&(deg)&&(MeV)&&(MeV) \\ \hline
145&0.078&$<$ 30&0.078(1)&2.0(1)&0.108(1)&2.5(2) \\ \hline
158&0.081&$<$ 30&0.082(2)&2.1(1)&0.119(4)&2.7(1) \\
&&$>$ 30&0.082(2)&2.1(1)&--&-- \\ \hline
170&0.084&$<$ 30&0.085(1)&2.3(2)&0.122(2)&2.8(1) \\
&&$>$ 30&0.085(1)&2.3(1)&--&-- \\ \hline
180&0.087&$<$ 30&0.088(2)&2.4(1)&0.125(3)&2.9(2)  \\
&&$>$ 30&0.088(2)&2.4(1)&--&-- \\ \hline
200&0.091&$<$ 30&0.093(2)&2.5(1)&0.131(3)&3.3(1) \\
&&$>$ 30&0.093(2)&2.5(1)&--&-- \\ \hline \hline
\end{tabular} \end{center}
\label{tbl1}
\end{table}

\begin{table}  [h]
\caption{Anisotropy parameters for proton emitted at different eenrgies.}
\begin{tabular}{ccccc} \hline \hline
Reaction&Energy (MeV)&$\beta_{fitted}$&$\beta_{j_{av}}$&$\beta_o$ \\ \hline
Ne + C&145&1.854 $\pm$ 0.504&1.641&5.678  \\
&158&1.188 $\pm$ 0.323&1.279&5.408  \\
&170&0.947 $\pm$ 0.163&1.041&4.937  \\
&180&0.937 $\pm$ 0.108&0.939&5.116  \\
&200&0.676 $\pm$ 0.164&0.844&4.911  \\ \hline
Ne + Al&158&3.861 $\pm$ 0.183&3.663&5.964 \\ \hline  \hline
\end{tabular}
\label{tbl2}
\end{table}

\newpage

Figures

\end{document}